\documentclass[12pt]{iopart}
\usepackage{iopams}
\usepackage{epsfig}
\usepackage{ulem}

\usepackage{bm}

\begin{document}

\title{Low-temperature spin dynamics of a valence bond glass in
  Ba$_2$YMoO$_6$.}

\author{M~A~de~Vries$^1$, J~O~Piatek$^2$, M~Misek$^3$, J~S~Lord$^4$,
  H~M~R\o{}nnow$^2$, J-W~G~Bos$^5$\eads{\mailto{m.a.devries@ed.ac.uk}}}
\address{$^1$ School of Chemistry, The University of Edinburgh, Edinburgh EH9 3JJ, UK.}
\address{$^2$ Laboratory for Quantum Magnetism, \'Ecole Polytechnique
  F\'ed\'erale de Lausanne (EPFL), Switzerland.}
\address{$^3$ School of Physics, The University of Edinburgh,
  Edinburgh, EH9 3JZ, UK.}
\address{$^4$ ISIS Facility, Rutherford Appleton Laboratory, Chilton,
  Didcot, Oxon OX11 OQX, UK.}
\address{$^5$ 
Institute of Chemical Sciences and Centre for Advanced Energy Storage
and Recovery, Heriot-Watt University, Edinburgh, EH14 4AS, UK}

\date{\today}

\begin{abstract}
We carried out AC magnetic susceptibility measurements and muon spin
relaxation spectroscopy on the cubic double perovskite Ba$_2$YMoO$_6$,
down to 50 mK. Below $\sim 1$~K the muon relaxation is typical of a
magnetic insulator with a spin-liquid type ground state, i.e. without
broken symmetries or frozen moments. However, the AC susceptibility
revealed a dilute-spin-glass like transition below $\sim 1$~K.
Antiferromagnetically coupled Mo$^{5+}$ $4d^1$ electrons in triply
degenerate $t_{2g}$ orbitals are in this material arranged in
a geometrically frustrated fcc lattice. Bulk magnetic susceptibility
data has previously been interpreted in terms of a freezing to a
heterogeneous state with non-magnetic sites where $4d^1$ electrons
have paired in spin-singlets dimers, and residual unpaired Mo$^{5+}$
$4d^1$ electron spins. Based on the magnetic heat capacity data it has been
suggested that this heterogeneity is the result of kinetic constraints
intrinsic to the physics of the pure system (possibly due to
topological overprotection), leading to a self-induced glass of
valence bonds between neighbouring $4d^1$ electrons.  The $\mu$SR
relaxation unambiguously points to a
heterogeneous state with a 
static arrangement of unpaired electrons in a background of
(valence bond) dimers between the majority of Mo$^{5+}$ $4d$
electrons. The AC susceptibility data indicate that the residual
magnetic moments freeze into a dilute-spin-glass-like state. This is
in apparent contradiction with the muon-spin decoupling at 50~mK in
fields up to 200 mT, which indicates that, remarkably, the time scale
of the field fluctuations from the residual moments is $\sim
5$~ns. Comparable behaviour has been observed in other geometrically
frustrated magnets with spin-liquid-like behaviour and the
implications of our observations on Ba$_2$YMoO$_6$ are discussed in
this context. 
\end{abstract}

\submitto{\NJP}

\maketitle

\section{Introduction}
Geometrical frustration of the exchange interactions in
antiferromagnetic (Mott) insulators with magnetic topologies based on
triangles and tetrahedra can fully suppress the spontaneous symmetry
breaking to the common long-range ordered antiferromagnetic (N\'eel)
state~\cite{Anderson:87, Neel:36}. Quantum-ordered spin-liquid states
are widely expected to occur in such systems. These could be gapless,
with emergent Fermionic quasiparticles called spinons with $S=1/2$ and
no charge or gapped (topologically-ordered)~\cite{Balents:10,
  Yan:11}. A number of materials with a spin-liquid-like absence of
frozen moments have now been identified~\cite{Shimizu:03, Mendels:07,
  deVries:09prl, Nakatsuji:12, Cheng:11}. These are without exception
gapless, often still display some weak hysteretic behaviour below 2~K,
and have only short-ranged dynamic magnetic correlations~\cite{deVries:09prl,
  Carlo:11,  Nakatsuji:12} which is contrary to theoretical
expectations of gapless spin liquids. 

While these short-ranged correlations could in some cases be due to
disorder~\cite{deVries:09prl, Nakatsuji:12}, another plausible
explanation is the presence of kinetic constraints towards freezing
into the (quantum) ground state \cite{Chandra:93, Chamon:05,
  Nussinov:07, Markland:11, Castelnovo:11, Olmos:12, Cepas:12}. It has
been shown that even where quantum fluctuations are so strong that no
local order parameters exist, as in some geometrically frustrated
quantum magnets, these quantum fluctuations are not always effective
for equilibration to its true quantum ground state~\cite{Chamon:05,
  Nussinov:07, Markland:11, Castelnovo:11, Cepas:12, Olmos:12}. When
the width of the tunnelling barriers becomes  comparable to the system
size, out-of-equilibrium quantum-glassy phases can
stabilize~\cite{Chamon:05, Nussinov:07, Markland:11, Castelnovo:11, Olmos:12}.
Quantum spin systems with strictly local interactions, with local
degrees of freedom coupled to a heat bath (Mott insulators), also
studied in the context of fault-tolerant quantum
computation~\cite{Kitaev:03}, are ideal model systems to study this
quantum glassiness~\cite{Chamon:05}. A particularly intriguing and
far-reaching possibility is that there are kinetic constraints due to
``topological overprotection''~\cite{Chamon:05, Castelnovo:11}. The
topological order~\cite{Wen:91} in a gapped quantum (spin liquid)
state is known to lend the ground state a certain rigidity and
``protect'' the quantum wavefunction against decoherence and material
imperfections~\cite{Ioffe:02, Kitaev:03}. An infinite number of local
operations would be required to mess up the macroscopic entanglement
of this quantum ordering. In Mott insulators this concept can be
turned on its head; here the physics is local~\cite{Zaanen:11} and
hence the only route to access any  topologically-ordered states in
the first place, is via an infinite number of local operations. This
has been termed ``topological overprotection''~\cite{Chamon:05,
  Castelnovo:11}. Within this context we studied the cubic fcc lattice
antiferromagnet Ba$_2$YMoO$_6$, which has previously been
suggested~\cite{deVries:10, McLaughlin:10} to freeze into a valence
bond glass (VBG) state~\cite{Tarzia:08}. This is in  line with
theoretical predictions for a quantum magnet with an fcc lattice
predicting a ``strong glass'' state due to topological overprotection
\cite{Chamon:05,   Castelnovo:11}. However, the Hamiltonian of this
cubic perovskite with unquenched orbital degrees of freedom is
significantly more complex than that of the Heisenberg Hamiltonian, as
was recently pointed out~\cite{GangChen:10}. 

\begin{figure}
  \begin{center}
  \epsfig{file=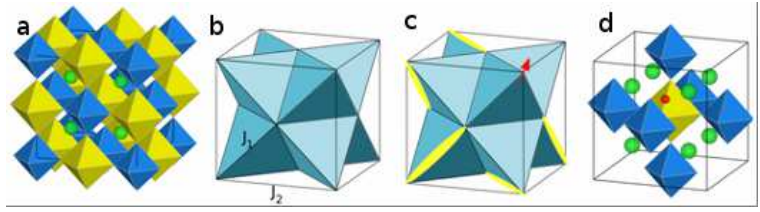, width=10cm}
  \caption
      {(Color online) (a) The Ba$_2$YMoO$_6$ double perovskite
        structure, showing the rocksalt ordering of MoO$_6$ octahedra
        (blue) and YO$_6$ octahedra (yellow) with Ba$^{2+}$ (green) at
        the perovskite A site. (b) The network of edge-sharing
        tetrahedra of the fcc antiferromagnet. (c) An illustration of
        the valence bond glass state, with the non-magnetic spin singlet
        dimers across random edges (yellow) and a single unpaired spin
        (red). (d) The muon stopping site (red) on an edge of the YO$_6$
        octahedron, surrounded by 6 MoO$_6$ octahedra.}
      \label{fig1}
      \end{center}
\end{figure}

Ba$_2$YMoO$_6$ is a double perovskite (of general formula
A$_2$BB$^{\prime}$O$_6$) where the two B-site cations have assumed
rocksalt order~\cite{Cussen:06} (Fig. 1a). Magic-angle spinning NMR
was reported to point to 3 \% cation disorder~\cite{Aharen:10} but in
neutron diffraction data of our sample the B-site ordering of the Mo/Y
cations refines to 100(1)\% 15, as expected for the large difference
in ionic radii of the Y$^{3+}$ and Mo$^{5+}$ of 0.29~\AA. The
Mo$^{5+}$ cations have a single unpaired electron in an octahedral
crystal field and are separated from each other by 90$^{\circ}$
Mo-O-O-Mo bonds ($J_1$) along the edges of corner-sharing tetrahedra,
and 180$^{\circ}$ Mo-O-Y-O-Mo bonds ($J_2$) (Fig. 1b). The 12
near-neighbour ($J_1$) bonds and 6 further neighbour bonds per site
lead to a strong net  antiferromagnetic interaction, with a Weiss
temperature of -143(5)~K. Ba$_2$YMoO$_6$ remains cubic (space group
$\mathrm{FM}\bar{3}\mathrm{m}$) down to at least 2~K which means the
single $4d$ $t_{2g}$  electron is also orbitally degenerate. Due to
relatively strong spin-orbit coupling in the $4d$ shell this
degeneracy is expected to be partially lifted, leading to $J = 3/2$
quadruplets~\cite{GangChen:10}. It is noteworthy that of the related
compounds where the Y$^{3+}$ is replaced for a trivalent lanthanide
ion only in the case of Nd$^{3+}$ and Sm$^{3+}$ a Jahn-Teller
distortion is observed to a tetragonal unit cell, and in both cases
this distortion is accompanied by antiferromagnetic
ordering~\cite{Cussen:06}. The actual Hamiltonian that applies to
Ba$_2$YMoO$_6$ is not reproduced here for brevity. It contains
antiferromagnetic and ferromagnetic exchange interactions that depend
on the orientations of the occupied orbital states within the
degenerate $t_{2g}$ manifold, a quadrupolar interaction between
neighbouring orbitals and a spin-orbit coupling term which can be
projected-out by constraining the system to a low-energy subspace with
$J = 3/2$ moments~\cite{GangChen:10}. A (na\"{i}ve) estimate of the
orbital-dependent nearest-neighbour antiferromagnetic exchange
interaction strength $J_1$, assuming further-neighbour interactions
can be neglected, yields $|J_1| \sim  72$~K. 

Remarkably, the magnetic/electronic heat capacity of the single $4d$
$t_{2g}$ electron at the Mo site evidenced a pseudogap below $\sim
50$~K, gradually locking up most of the magnetic entropy of the $J =
3/2$ moments~\cite{deVries:10}. This ``strong glass''-like freezing
does however not translate to a significantly increased muon
relaxation down to 2~K, indicating that most of the Mo$^{5+}$ $4d$
electrons pair up with near-neighbour spins to form non-magnetic
spin-singlet- or valence bond dimers~\cite{deVries:10} (Fig.~1c). A
second Curie-Weiss regime is observed in the DC susceptibility below
25~K, with a Weiss temperature of $-2.6(2)$~K. The paramagnetic
susceptibility between 1.9 and 7~K was found to be equivalent to that
from 8(1)\% of the Mo$^{5+}$ magnetic moments as measured at high
temperature (see supplementary information). These residual,
apparently weakly-coupled spins (sometimes referred to as ``orphan
spins'') are often observed in geometrically-frustrated
compounds~\cite{Schiffer:97}. In some cases this has been suggested to
be spins left-out from spin-singlet valence-bond dimerizations between
the majority of spins, leaving a small fraction of
magnetically-isolated and thus weakly-interacting unpaired-spin
moments~\cite{Uemura:94MuSCGO, deVries:10} (Fig.~1c). Hence,
``dangling spins'' might in this case be a better term. The aim of the
present study was to characterise the low-temperature spin-orbital
dynamics to test the VBG proposal.

\section{Results}
The polycrystalline sample was synthesised using a solid state
synthesis method as described in Ref.~\cite{deVries:10}.  AC magnetic
susceptibility measurements were carried out using an inductive
susceptometer from CMR-direct in an Oxford Instruments dilution fridge
(down to 50~mK) and in a $^3$He fridge (down to 300~mK). The
measurements used an AC field of 0.02 mT in a frequency range from
54.5 to 8900~Hz. The intensity of the AC susceptibility signal at all
frequencies was assumed to match the low-temperature DC susceptibility
obtained using a SQUID magnetometer [see supplementary
  information]. As shown in Fig. 2a,b frequency-dependent cusps are
observed around 600~mK in both the dispersive ($\chi ^{\prime}$) and
dissipative ($\chi ''$) parts of the dynamic susceptibility. The
frequency dependence of $T_{\textrm{max}}$ (Fig. 2c) exhibits
Arrhenius-activated behaviour (the straight line in Fig. 2c), with an
activation energy $E_a = 39(2)$~K and a trial frequency $f_0 \sim
1\times10^{29}$~Hz. Such a frequency dependence in the AC
susceptibility is a defining feature of a strong glass as well as spin
glasses~\cite{Cannella:72}. Figure 2d shows the effect of an
additional DC field on the AC susceptibility at the spin-glass
transition, revealing only a modest suppression of the AC
susceptibility in fields below 250~mT. The AC magnetic susceptibility
is a macroscopic property. We complemented this data with muon spin
relaxation spectroscopy which is a local probe of the spin dynamics in
the material, using the MUSR instrument at the ISIS facility,
UK. Figure 3a shows the  average time dependence of the spin
polarisation of muons implanted in a 3~g powder sample, for a
selection of temperatures between 50~mK and 120~K and in zero external
field. Figure 3b shows the field dependence of the muon relaxation at
50~mK, for (longitudinal) magnetic fields applied along the initial
muon spin polarisation direction. 

\begin{figure}
\begin{center}
\epsfig{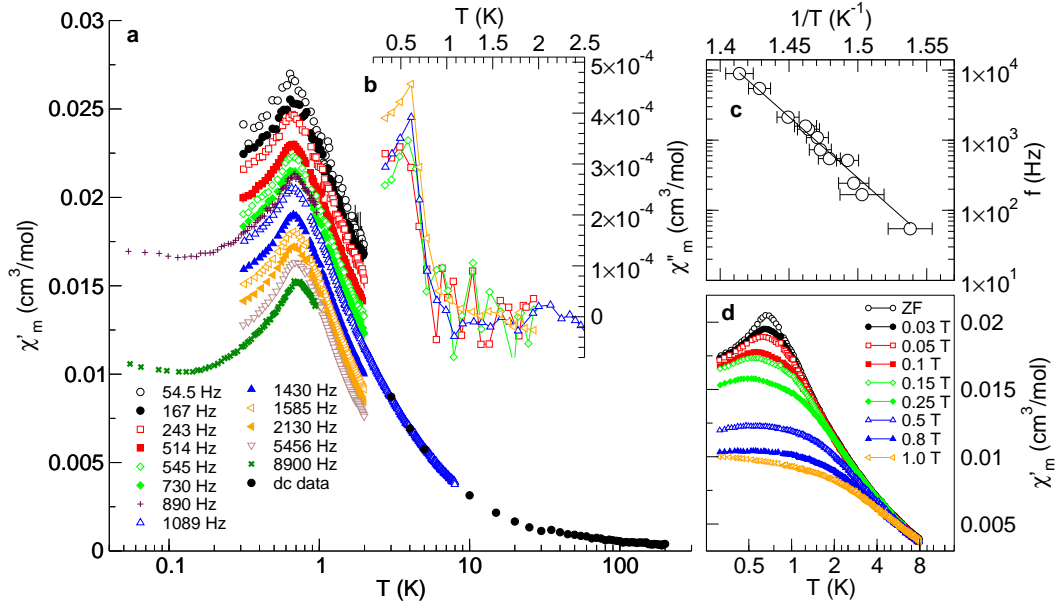}
\caption{(Color online) AC magnetic susceptibility measured in a
  dilution refrigerator. (a) The frequency dependence of dispersive
  (real) part of the AC susceptibility at low temperatures matched to
  the DC susceptibility. (b) The dissipative (imaginary) part of the
  AC susceptibility near the spin-glass transition. (c) The Arrhenius-
  activated frequency dependence of the temperature with maximum AC
  susceptibility ($T_{\textrm{max}}$) (see text) (d) The effect of DC
  fields up to 1 T on the AC susceptibility (at 1029~Hz) near the
  transition temperature.}
\end{center}
\end{figure}

At the time the muons are implanted in the sample ($t = 0$), all muon
spins point along the forward ($z$) direction. The muon spins will
then precess due to the local fields at the muon sites. In a material
with uniform magnetic order all muons will precess with the same
frequency, which translates to an oscillation of the muon polarisation
$P_z(t)$. In the case of static disordered moments a single damped
oscillation is observed equilibrating to 1/3rd of the initial
polarisation, corresponding to on average 1/3rd of the randomly
orientated magnetic moments that are aligned with the initial muon
spin polarisation and hence do not cause muon precession or
depolarisation. This type of relaxation is observed, for example,
below the spin-glass transition in the related compound
Sr$_2$MgReO$_6$ with Re$^{6+}$ $S = 1 / 2$~\cite{Wiebe:03}. The muon
relaxation in the presence of static but random fields can be
suppressed by the application of a longitudinal magnetic field
stronger than the local fields at the muon site. This enabled the
determination of the contribution to the muon relaxation from nuclear
spins $P_{nuc}(t)$ at the Y and at a fraction of the Ba and Mo sites
in Ba$_2$YMoO$_6$, which below $\sim 10$~K are static at the time
scale of the muon experiment. At 1.3~K, just above the spin-glass
transition observed in the AC susceptibility data, a 5~mT field was
sufficient to decouple the muon relaxation for the small nuclear
fields (typically up to 1~mT ), yielding the nuclear relaxation factor
$P_{nuc}(t)$ as shown in Fig. 3b. The observed nuclear relaxation is
as expected for the muon stopping site on the edges of the YO$_6$
octahedra~\cite{Cherry:95} (Fig. 3d, see supplementary information for
further details). No further change in the muon relaxation was
observed with the application of longitudinal fields up to 200~mT,
confirming that the remaining muon relaxation was due to larger
(non-frozen) electronic moments. Well into the paramagnetic state the
electron-spin fluctuations are usually too fast to be picked-up by
muons leading to negligible muon relaxation at 120~K (Fig. 3a). As
indicated by the heat capacity data, the entropy from most of the $J =
3/2$ moments is locked up below $\sim 40$~K. The very slow relaxation
at 2~K thus confirms that the majority of these magnetic moments must
have paired into non-magnetic spin-singlet or valece bond
dimers~\cite{deVries:10}. Across the spin-glass transition the muon
relaxation rate increases sharply but even at 400 mK, below which no
further increase in the muon relaxation (depolarization) is observed,
the relaxation rate remains remarkably low, even for a system with
dilute electronic spins. Amongst antiferromagnetic insulators without
a spin-gap, only in the spin-liquid compound herbertsmithite is a
slower muon relaxation observed at 50 mK~\cite{Mendels:07}. If this
weak relaxation were due to static random moments, then the fields
from these moments at the muon site would have to be as small as $\sim
0.6$~mT, so that the muon relaxation should be fully decoupled in a
5~mT field. As is clear from Fig. 3b, a 5~mT field only leads to a
small reduction of the relaxation, the difference with the zero-field
relaxation being given approximately by $P_{nuc}(t)$. A clear muon
relaxation signal persists in fields up to 200~mT, consistent with the
presence of fluctuating electronic magnetic moments.

\begin{figure}
  \begin{center}
    \epsfig{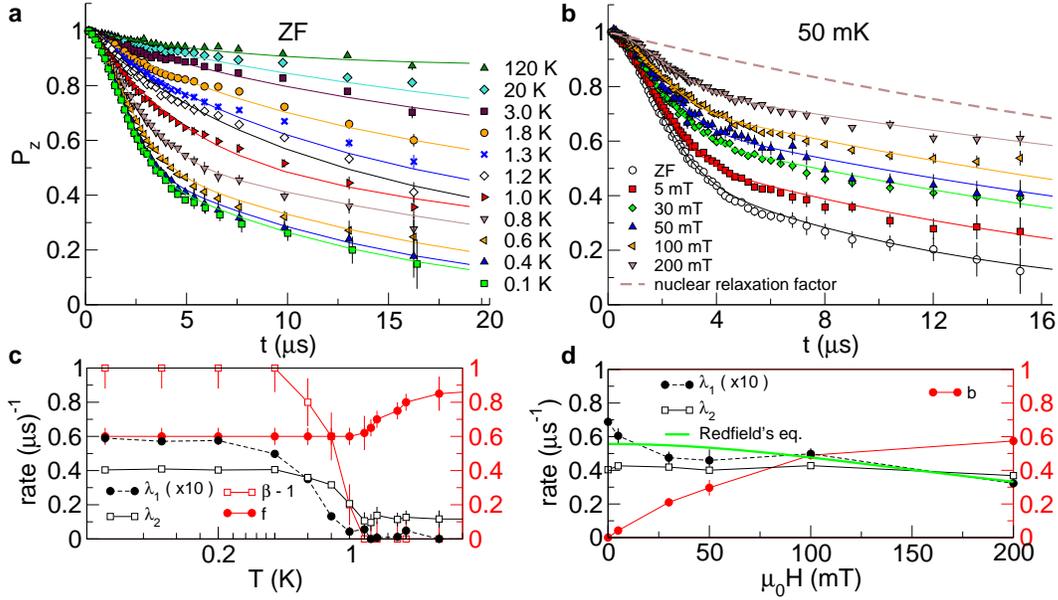}
    \caption{(Color online) Muon spin relaxation data obtained at MUSR,
      ISIS. (a) The 
      temperature dependence of the zero- field muon relaxation ($P_z(t)$)
      between 50~mK and 120~K. (b) The field dependence of the muon
      relaxation at base temperature (50~mK) and the nuclear relaxation
      factor which was confirmed to be constant below ~ 1.3~Κ (See text
      and supplementary material). (c) The fit parameters for fits of
      Eq. 1 to the zero-field data as a function of temperature, taking
      into account the weak nuclear relaxation. (d) The fit parameters for
      fits of Eq. 1 to the 50~mK muon relaxation in fields up to
      200~mT. The fluctuation rate was determined from the field
      dependence of $\lambda _1$ using Redfield’s equation (solid
      green line) (see text and supplementary information).}
  \end{center}
\end{figure}

The muon relaxation $P_z(t)$ is determined by the magnitude, variation
and fluctuation of local fields inside a sample, and carefully
analysing the shape of the muon relaxation curve can allow to
distinguish different scenarios. Here we found that the best fitting
scenario is with two different muon environments, designated E1 and
E2, with an exponential relaxation in E1 (arising from spin
fluctuations with a rate $\lambda _1$ that will be estimated later), and a
phenomenological modified-exponential decay with $1 \leqslant \beta
\leqslant 2$ (for E2);
\begin{equation}
 P_z (t) = P_{\mathrm{nuc}}(t)\left[ f e^{ - \lambda _1 t} +(1-f)\left(
   e^{ - (\lambda _2 t )^\beta}+b \right)\right],
\end{equation} 
where $P_{\mathrm{nuc}}(t) = 1$ for longitudinal fields $\geqslant
5$~mT, $f$ is the fraction of  muons in E1 and the constant background
$b = 0$ in zero-field. The results of the fits to the data down to 50
mK and in fields up to 200~mT are shown in Fig. 3c and d,
respectively. The background contribution $b$ in E2 is constrained by
the condition that with the fields up to 200 mT applied here the
fraction of muons in E1 and E2 should be constant, as implied by the
AC susceptibility data in Fig. 2d. The increase of this
  phenomenological parameter $b$ with increasing fields accounts for
  the gradual decoupling of the muons  from the electron spins in
  E2. Just above the spin glass transition, at 1.3 K, the stretching
exponent $\beta = 1$ and $\lambda _1 = 0$. This implies that a major
fraction $f$ of the muons is located in a non-magnetic environment
(E1), while the remaining muons ($\sim 30$\% at 1.3~K) undergo an
exponential relaxation due to surrounding fluctuating spins (E2). On
cooling through the spin-glass transition the depolarisation of the
muons in E2 changes sharply from exponential to Gaussian, as
previously observed in the $S = 3/2$ kagome antiferromagnet
SrCr$_8$Ga$_4$O$_{19}$~\cite{Uemura:94MuSCGO}, and the fraction of
muons in E2 increases to 40(5)\%. At the lowest temperatures accessed
the muons in E1 also exhibit a weak exponential relaxation (Fig. 3c).
At base temperature E1 and E2 correspond to muons surrounded by
respectively 0 and 1 unpaired electron on the 6 Mo$^{5+}$ ions
surrounding the muon site. Note, however, that in E1 muons will
  still be affected by unpaired electrons at more distant sites,
  especially at low temperatures. ``Non-magnetic'' (E1) and magnetic
(E2) environments were also evident from $^{89}Y$ NMR~\cite{Aharen:10}
with a ratio of approximately 50:50, comparable to our results
(Fig. 3c). This confirms that 1) the muon stopping site is near the
Y$^{3+}$ and 2) that the muon does not perturb the low-temperature
dilute spin-glass state significantly. Because each muon  on the Y
site is surrounded by 6 Mo$^{5+}$ ions, the fraction of muons in a
non-magnetic environment of 0.60(5) implies that at low temperature (
$< 10$~K) a fraction of 0.91(1) of the Mo$^{5+}$ sites [ $(0.91)^6 =
  0.6$] are non-magnetic. Hence, only $\sim 9(1)$\% of the Mo$^{5+}$
carry a net magnetic moment at low temperature.  Below 400 mK no
further increase of the muon relaxation rates is observed down to 50
mK, indicating that thermal fluctuations do not play a major role at
these temperatures. The spin fluctuation rate $\tau$ and magnitude
$\Delta$ can in ideal cases be determined from the field dependence of
the exponential muon relaxation rate $\lambda _1$ using Redfield’s
formula~\cite{Baker:11} (see supplementary information). The field
dependence of the exponential relaxation rate $\lambda _1$ in E1 at 50
mK as obtained from the data in Fig. 3b was fitted with Redfield’s
formula (the green line in Fig. 3d, see also supplementary material),
yielding a fluctuation time scale of 5(2)~ns, with a field
distribution $\Delta$ of 2.2(1)~mT. This field-strength is as expected
for muons in E1, without directly neighbouring unpaired spins. These
muons probe the average fluctuating magnetic field from many distant
unpaired electrons.

\section{Discussion}
Our analysis of the muon data rules out any major rearrangement of the
unpaired spins over the available Mo sites within the time scale of
the muon experiment of 16~$\mu$s. Dynamical models of fluctuating
valence bonds, where the orphan- or dangling spins are mobile, as
proposed for SrCr$_8$Ga$_4$O$_{19}$~\cite{Uemura:94MuSCGO} have been
explored but can in the case of  Ba$_2$YMoO$_6$ be ruled out because
of the much larger fraction of dangling spins, while at the same time
the muon relaxation is more than an order of magnitude slower than for
SrCr$_8$Ga$_4$O$_{19}$~\cite{Uemura:94MuSCGO}. Such models predict a
fast $t = 0$ muon relaxation for a significant fraction of the muons,
which is not observed; the $t = 0$ asymmetry at 50~mK and 120~K are
comparable. Hence, the low temperature state in Ba$_2$YMoO$_6$ is
an inhomogenous state without any sign of local order
parameters down to 50 mK. This state contains non-magnetic sites where
the neighbouring Mo$^{5+}$ $4d^1$ electrons must have paired up in
non-magnetic spin-singlets, and residual unpaired spins. A VBG state,
as previously proposed~\cite{deVries:10}, provides a natural
explanation for this inhomogeous state. However, a larger than
previously estimated~\cite{Aharen:10, deVries:10} amount of Mo/Y
antisite disorder would also explain the presence of residual unpaired
spins at fixed locations. Hence, of all data so far available on
Ba$_2$YMoO$_6$ the strongest evidence in favour of a VBG is the heat
capacity~\cite{deVries:10}. As pointed out earlier, the pseudogap in
the heat capacity~\cite{deVries:10} below 50~K (also visible in the
neutron spectra~\cite{Carlo:11}) points to a  gradual, strong-glass,
freezing at temperatures comparable to the magnetic exchange
interaction $|J_1|$ of $\sim 73$~K. Hence, it must be the readiness of
each $4d$ electron to randomly form a near-neighbour valence bond with
one of its 12 near neighbours at high temperature, that prevents
freezing into a (quantum-) ordered structure at low temperature. This
type of behaviour is also characteristic of structural glass formers.
It is also thought that disorder, even a very small amount, can
stabilise a VBG state~\cite{Tarzia:08,Singh:10}, although  disorder is
also widely known to reduce certain kinetic constraints towards
nucleation of the thermodynamic state. An intriguing question that
therefore remains is whether the thermodynamic state would be a
topologically ordered state, in which case the kinetic constraints
stabilising the VBG are due to topological overprotection, or an
orbitally ordered state leading to a structural distortion from cubic
to tetragonal symmetry at low temperature. It should in this respect
be noted that the latter is observed in the closely related materials
Ba$_2$NdMoO$_6$ and Ba$_2$EuMoO$_6$, while analogues with Gd, Dy, Er,
Yb have a magnetic susceptibility similar to that of
Ba$_2$YMoO$_6$~\cite{Cussen:06}. 

We now turn to the observations on the $\sim 8$\% of residual unpaired
spins, that account for the low-temperature magnetic
susceptibility. The Arrhenius-activated behaviour in the
low-temperature AC susceptibility indicates that these  spins freeze
into a dilute-spin-glass-like state, with thermally-activated
collective dynamics at the macroscopic scale. This implies that these
spins interact via a weak longer-distance exchange
coupling (dipolar  couplings would lead to freezing at still lower
temperatures), and are randomly distributed over the Mo sites. It is
surprising that even for muons directly neighbouring these unpaired
spins the relaxation down to 50~mK remains extremely slow; if the
residual spins giving rise to the spin-glass response in the AC
susceptibility were static at the time scale of the muon experiment,
this would lead to fast relaxation of 2/3rd of the implanted muons
well within 1~$\mu$s, as observed in dilute
spin-glasses~\cite{Dodds:83}. Less extreme examples where these
apparently disparate behaviours coexist are known amongst
geometrically frustrated magnets with ground states sometimes
described as ``quantum disordered'' ~\cite{Schiffer:97,
  Uemura:94MuSCGO}. The slow relaxation of the muons can only in part
be explained by the large distance between the muon site and the
nearest spin moments. A full understanding of this common behaviour in
low-spin frustrated magnets is still outstanding. It is possible that
the implanted muons perturb the fragile low-temperature spin-glass
state, but this raises the question why below $\sim 10$~K the nuclear
spins usually remain static on implantation of a
muon. Zero-temperature fluctuations might also arise due to
decoherence that affects ungapped quantum states. As a result the
quantum glass state is best represented by a density matrix and not as
a single wave function~\cite{Chamon:05}. This is however unlikely to
account for the difference in behaviour at microscopic and macroscopic
scales. A third explanation that should be considered is that the
explanation lies exactly in the difference of length scales; that the
spin-glass freezing observed in the (macroscopic) magnetic
susceptibility is truly an emergent property, while at microscopic
length scales the spin dynamics is dominated by quantum
fluctuations. These quantum fluctuations are in this case ineffective
in equilibrating the system to its thermodynamic (quantum-) ordered
ground state~\cite{Chamon:05, Castelnovo:11, Markland:11}. A time
resolved coherent x-ray diffraction experiment on out-of-equilibrium
states with local order parameters does suggest~\cite{Shpyrko:07} that
quantum fluctuations can at microscopic length scales occur at ns time
scales, as also suggested by our data. 

\section{Conclusion}
Our results confirm that the magnetic low-temperature state in
Ba$_2$YMoO$_6$ without broken symmetries in the magnetic degrees of
freedom is an inhomogeneous state. The lattice positions of the
  residual moments could be dynamic but at a time scale that is slow
compared to the time scale of the muon experiment of 10 $\mu$s, with
the majority of spins paired into valence bonds. That this is a
valence bond glass rather than a quantum spin liquid is in this case
concluded from the pseudogapped electronic/magnetic heat
capacity~\cite{deVries:10}. The $\mu$SR experiment carried out here
does not distinguish between a spin liquid and a valence bond glass;
the relaxation is very similar to that reported for many materials
presumed to have spin-liquid-like ground
states~\cite{Mendels:07}. However, they do confirm, for the first
time, that the ground state is heterogeneous and without any broken
symmetries down to 50~mK. 

These results highlights the importance of the electronic/magnetic
heat capacity to characterise materials with spin-liquid like ground
states. In the absence of such data, as for the
well-known kagome antiferromagnet spin liquid material
herbertsmithite, it remains difficult to say whether spin liquid or a
valence bond glass states stabilize. The possibility of a VBG in
herbertsmithite due to a small amount of
disorder~\cite{deVries:10nat} was put forward~\cite{Singh:10}
following neutron spectroscopy data displaying below 120 K temperature
independent short-ranged antiferromagnetic dynamic
correlations~\cite{deVries:09prl}, very much like the neutron spectrum
observed for Ba$_2$YMoO$_6$~\cite{Carlo:11}.  

\ack
We would like to thank the Science and Technology
Facilities Council for the provision of muon beamtime and Dr. P. King
(ISIS) for assistance with the muon spectroscopy measurements at
MUSR. We gratefully acknowledge discussions with
Dr. P.~J. Camp (Edinburgh), Dr. A.~C.  McLaughlin (Aberdeen) and
Dr. C. Castelnovo (Cambridge) and helpful comments by Dr. J. Zaanen
(Leiden). Part of this work was supported by funding from the
Swiss National Science Foundation, MaNEP and Sinergia network MPBH.

\providecommand{\newblock}{}

\end{document}